\documentclass[twocolumn,showpacs,amsmath,amssymb,aps,pra,floatfix,nofootinbib,superscriptaddress]{revtex4}

\usepackage{graphicx,graphics}% Include figure files
\usepackage{bm}% bold math
\begin{document}

\title{Quantum dense coding by spatial state entanglement}
\author{O. Akhavan}
\email{ akhavan@mehr.sharif.edu} \affiliation{Department of
Physics, Sharif University of Technology, P.O. Box 11365-9161,
Tehran, Iran} \affiliation{ Institute for Studies in Theoretical
Physics and Mathematics (IPM), P.O. Box 19395-5531, Tehran, Iran}

\author{ A.T. Rezakhani }
\email{ tayefehr@mehr.sharif.edu} \affiliation{Department of
Physics, Sharif University of Technology, P.O. Box 11365-9161,
Tehran, Iran}
\author{ M. Golshani}
\email{golshani@ihcs.ac.ir}
\affiliation{Department of Physics,
Sharif University of Technology, P.O. Box 11365-9161, Tehran,
Iran} \affiliation{ Institute for Studies in Theoretical Physics
and Mathematics (IPM), P.O. Box 19395-5531, Tehran, Iran}
\begin{abstract}
We have presented a theoretical extended version of dense coding
protocol using entangled position state of two particles shared
between two parties. A representation of Bell states and the
required unitary operators are shown utilizing symmetric
normalized Hadamard matrices.  In addition, some explicit and
conceivable forms for the unitary operators are presented by using
some introduced basic operators.  It is shown that, the proposed
version is logarithmically efficient than some other multi-qubit
dense coding protocols.
\end{abstract}
\date{\today}
\pacs{03.67.Hk, 89.70.+c} \maketitle

%%%%%%%%%%%%%%%%%%%%%%%%%%%%%%%%%%%%%%%%%%%%%%%%%%%%%%%%%%%%%%%%%%%%%%%%%%%%%%%%%%%%%
The quantum entanglement property is providing new methods of
information transfer, in some cases much more powerful than their
classical counterparts.  In quantum information theory,
entanglement, as a key concept is used for a wide range of
applications, such as dense coding {\cite{BW92}}, teleportation
{\cite{BB93}}, secret sharing {\cite{HBB99}} and key distribution
{\cite{IEEE,Deutsch}}. To see more about the mentioned topics and
the efforts done on them, both theoretically and experimentally,
one can refer to {\cite{BEZ,Cab}} and references therein.

Quantum dense coding protocol proposed originally by Bennett and
Wiesner {\cite{BW92}} in 1992.  The protocol describes a way to
transmit two bits of classical information through manipulation of
only one of the entangled pair of spin-$\frac{1}{2}$ particles,
while each of the pair individually could carry only one bit of
classical information.  The first experimental realization of
dense coding has been reported by Mattle {\it et al.} {\cite{Mat}}
in 1996.

There have been attempts to generalize dense coding protocol to
achieve higher channel capacity. The first proposition in this
regard was due to the original reference of dense coding
{\cite{BW92}} by using a pair of $n$-state particles prepared in a
completely entangled state (instead of an EPR spin pair in a
singlet state) to encode $n^{2}$ values.  In practice, there might
be some limitations for finding $n$-state particles with high
$n$'s and controlling them.  Although very recently, some progress
have been made in this direction {\cite{Molina}}, but still it is
worth, both theoretically and experimentally, examining other
alternatives to achieve this goal. For example, Bose {\it{et al.}}
{\cite{Bose}} studied $\textsf{N}$ pairwise entangled states in
which each party gets one particle except Bob with ${\textsf{N}}$
qubits. Also in a more efficient scheme, they considered
${\textsf{N}}+1$ parties sharing maximally entangled qubits in a
way that each party, including Bob, possesses one qubit.  In
addition, some recent attempts on generalization of quantum dense
coding can be found in {\cite{Liu,Grudka,Werner,Gorbachev}}.  In
an elegant alternative, Vaidman {\cite{Vaid}} proposed a method
for utilizing canonical continuous variables $x$ (position
variable) and $p$ (linear momentum variable) to perform a quantum
communication. After that, Braunstein and Kimble {\cite{Kim}}
presented a typical realization for continuous variable dense
coding using the quadrature amplitudes of the electromagnetic
fields in which the mean photon number in each channel should be
considered very large.  Recently, in another way, some effort has
been also done on experimental realization of continuous variables
dense coding {\cite{Li}}.

In this work, we have theoretically presented another extended
version of dense coding which uses discrete spatial variables
along with only two entangled particles.  In this regard, all
necessary Bell states and their corresponding unitary operators
are presented to encode and decode information.  This version at
large $N$'s can be considered as a conceivable scheme for
Vaidman's idea {\cite{Vaid}} except that we have considered just
the position variable (not both canonical variables $x$ and $p$)
for communication.  Finally, the efficiency of our scheme is
compared with some other known ones.  Furthermore, according to
Werner's general classification on dense coding and teleportation
schemes \cite{Werner}, it is possible to find a teleportation
protocol corresponding to our dense coding scheme.

\begin{figure}[t]
\includegraphics[width=3.5cm,height=7.5cm,angle=-90]{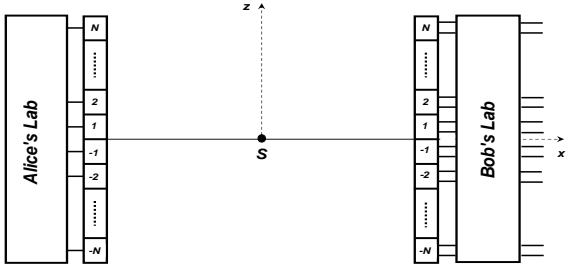}
\caption{Schematic of the proposed dense coding scheme.  The $2N$
receivers of each party are connected to its corresponding lab by
quantum channels.  The quantum channels can be connected together
in each lab using side quantum channels.  Alice's lab is equipped
with $O_{(k_{r},j)}$ encoder unitary operators, and Bob's lab
contains a decoder unitary operator like ${\cal H}$.  The Bell
state measurement is completed by just a simple position
measurement on Bob's outgoing channels.} \label{fig1}
\end{figure}

Consider an original EPR source which emits isotropically a pair
of identical (fermionic or bosonic) particles with vanishing total
linear momentum in a two dimensional space.  The similar EPR
source is also used elsewhere to clarify some discussions on the
foundations of quantum mechanics {\cite{Gol}}. The source $S$ is
placed exactly in the middle of the two parties, ``Alice" and
``Bob", where each one has an array of receivers aligned on a
vertical line. The receivers just {\em{receive}} the particles and
do not perform any destructive measurement. Total number of the
receivers of each party is considered to be $2N$. Since Alice and
Bob are assumed to be very far from each other and the source is
considered isotropic, there is an equal probability for every
receiver to obtain one of the emitted particles.  We label the
receivers placed at the upper (lower) part of the $x$-axis with
positive (negative) integers; 1, 2, \ldots, $N$ (-1, -2, \ldots,
$-N$). Figure~{\ref{fig1}} shows an illustration of this scheme.

 Now, the position state of the {\em system} (the arrays + the source)
can be written in the form
\begin{eqnarray}{\label{position}}
&|\psi_{1,2}\rangle=\frac{1}{\sqrt{2N}}\sum_{n=1}^{N}[|n,-n\rangle\pm|-n,n\rangle]
\end{eqnarray}
where the subscripts $1$ and $2$ are related to $\pm$ signs,
respectively, and $n$ refers to label of the receivers.  In
addition, the order in writing the state is according to direct
product of Alice's state and Bob's.  The signs $\pm$ in
$|\psi_{1,2}\rangle$ indicate the symmetry and anti-symmetry
property of the position state with respect to the particles
exchange. Without loss of generality, we assume bosonic property
for our system.  In general, the entangled state
({\ref{position}}) can be considered as a member of a larger
family (with a bit different notation), that is,
\begin{eqnarray}\label{firstfamily}
&|\psi_{(1,j)}\rangle=\frac{1}{\sqrt{2N}}\sum_{n=1}^{N}[h_{j,2n-1}|n,-n\rangle+h_{j,2n}|-n,n\rangle]\nonumber\\
&1\leq j\leq 2N
\end{eqnarray}
in which $h_{i,j}=\sqrt{2N}[{\bf{H}}]_{i,j}$, and $\bf{H}$ is a
$2N$-dimensional {\em{normalized symmetric}} Hadamard matrix which
satisfies the property ${\bf H}^{2}=I$.  In addition, the above
entangled states can be generalized to a more complete set of
orthonormal and maximally entangled states, which can be defined
as
\begin{eqnarray}\label{totPsi}
& \hskip -60mm |\psi_{(k_{r},j)}\rangle=\nonumber\\&\frac{1}{\sqrt{2N}}\sum_{n=1}^{N}[h_{j,2n-1}|n,f_{k_{r}}(n)\rangle+h_{j,2n}|-n,-f_{k_{r}}(n)\rangle]\nonumber\\
&1\leq k\leq N, 1\leq j\leq 2N
\end{eqnarray}
%\vspace{11mm}
%
%\begin{widetext}
%\begin{center}
where $k_{r}$ and $j$ are called family and member indices,
respectively.  Furthermore, we have adopted the conventions
$h_{j,i|j\leq0 ~{\text{or}}~ i\leq0}=0$ and
$f_{k_{r}}(n)=r(n+k-1)_{{\text{mod}}(N)}$ where $r=\pm 1$.  Now,
the states ({\ref{totPsi}}) form {\em{Bell bases}} for our dense
coding scheme. In fact, it is straightforward to check that these
states have the properties
\begin{eqnarray}{\label{orthonormality}}
&\langle\psi_{(k_{r},j)}|\psi_{(k'_{r'},j')}\rangle=\delta_{j,j'}\delta_{k,k'}\delta_{r,r'}\\
&{\text{tr}}_{1(2)}(|\psi_{(k_{r},j)}\rangle\langle\psi_{(k_{r},j)}|)={\frac{1}{2N}I_{1(2)}}.
\end{eqnarray}
Here it should be noted that, the order of a typical real Hadamard
matrix can be 1, 2 and $4k$ where $k$ is a positive integer
{\cite{Hadamard}}.  Physically it means that in other cases one
cannot make enough necessary entangled orthonormal states to
perform an efficient dense coding.

For allowed $N$'s, Alice should find unitary encoding operators
which transform the state describing the system in
Eq.~(\ref{position}), $|\psi_{1}\rangle$, as a Bell state with
$j=1$ (and $k=1,$ $r=-1$) into the others given in
Eq.~(\ref{totPsi}). The representation of $4N^2$ suitable unitary
operators for this task is
\begin{eqnarray}{\label{O}}
&\hskip -6.5cm O_{(k_{r},j)}=\nonumber\\
&\sum_{n=1}^{N}[h_{j,2n-1}|n\rangle\langle
f_{k_{r}}(n)|+h_{j,2n}|-n\rangle\langle -f_{k_{r}}(n)|]
\end{eqnarray}
which acts as follows
\begin{eqnarray}\label{opsi}
O_{(k_{r},j)}|\psi_{(k'_{r'},j'=1)}\rangle=|\psi_{(k''_{r''},j)}\rangle
\end{eqnarray}
in which
\begin{eqnarray}
k''_{r''}:=(k+k'-1)_{rr'}{{\text{mod}}(N)}. {\label{k}}
\end{eqnarray}

It is a relevant question how to implement $O_{(k_{r},j)}$
operators in practice. So here, we concisely propose one way to
perform this task using some basic and conceivable operators. For
example, the same as Pauli's operator in the spin-$\frac{1}{2}$
space, suitable basic operators in the position space can be
considered to be
\begin{eqnarray}{\label{Nn}}
&&N_{n}|\pm n\rangle=\pm|\pm n\rangle \\
&&P_{n}|\pm n\rangle=|\mp n\rangle
\end{eqnarray}
where the subscript $n$ means that the operator acts as a local
gate on the $\pm n$-th channels and as identity on the other
channels. The latter operator, $P_n$, is defined to relate the two
groups of receivers in the upper and lower halves of the $x$-axis.
We also need to define a ladder operator $L_{+}$ to relate the
receivers located on each half of the $x$-axis, so that
%\begin{subequations}
\begin{eqnarray}{\label{L+}}
&&L_{+}|\pm n\rangle=|\pm(n+1)\rangle_{{\text{mod}}(N)}.
\end{eqnarray}
%\end{subequations}

It is easy to understand that the main task of the basic operators
({\ref{Nn}}-\ref{L+}) is really displacement of a channel or
equivalently displacement of a particle from a channel to the
other.  This task can be performed using, for example, some side
quantum channels which connect the main channels together in a
suitable way.  So, if our introduced channels in the scheme are
considered, for instance, superconductor wires containing an
entangled current according to (\ref{position}), then these basic
operators can be conceived to be realized by using switching
process in the superconductor circuits. In fact, when a basic gate
is OFF, its main channel(s) is (are) superconducting and the side
channels are not superconducting, and if the gate is ON then the
main channel(s) is (are) not superconducting and the suitable and
corresponding side channel(s) is (are) superconducting.  However,
full realization of this proposition is left as an experimental
challenge.

Now using the above basic gates, we have shown that one can
construct operators such that transform any member of a family to
any other member of the other ones.  Here, our construction
mechanism is to find a set of operators, $O_j$, to transform the
$j$ label of the Bell states of any given family. Next, we
introduce another set of operators, $F_{k_{r}}$, to transform $k$
label of the Bell states of any given member. Thus, our desired
total operators are
\begin{eqnarray}{\label{totalO}}
O_{(k_{r},j)}=F_{k_{r}}O_{j}.
\end{eqnarray}
The explicit form of the $O_j$ operators can be constructed using
a normalized symmetric Hadamard matrix as follows
\begin{eqnarray}\label{O_j}
&O_j=\prod_{i=1}^{N}(P_{i}N_{i}P_{i})^{(h_{a,2i-1}-h_{j,2i})/2}N_{i}^{(h_{a,2i}-h_{j,2i})/2}\nonumber\\
&1 \leq j,a\leq 2N
\end{eqnarray}
where $a$ is an arbitrary fixed positive integer. One can easily
check that, the operator $O_j$ has the property $
O_{j}|\psi_{(k_{r},j')}\rangle=|\psi_{(k_{r},j'')}\rangle$ where
$h_{j'',i}=h_{j,i}h_{j',i}$. Furthermore, the explicit form for
$F_{k_{r}}$ can be considered as
\begin{eqnarray}\label{F_k}
&F_{k_{r}}=L_{+}^{(1-k)} \prod_{i=1}^{N}P_{i}^{(1-r)/2}.
\end{eqnarray}
It can be seen that they transform families to each other
according to
$F_{k_{r}}|\psi_{(k'_{r'},j')}\rangle=|\psi_{(k''_{r''},j')}\rangle$
where $k''_{r''}$ satisfies the rule in Eq.~(\ref{k}).

Now, Bob is ready to perform a Bell state measurement (BSM) on his
particle and the processed particle which is sent to him by Alice.
To do a BSM, Bob needs apply a {\it grand} unitary and Hermitian
operator on all the channels in his lab which should be followed
by a position state measurement on his outgoing channels.  To
simplify the representation of the grand operator we first move
into another set of Bell states which have the compact form as
$|\psi
'_{(k_{r},j)}\rangle=\frac{1}{\sqrt{2N}}\sum_{n=1}^{2N}h_{j,n}|n,f'_{k_{r}}(n)\rangle$
where $f'_{k_{r}}(j)=r(j+k-1)_{{\text{mod}}(2N)}$.  These states
can be obtained from the firstly defined states,
Eq.~(\ref{totPsi}), by application of some local unitary operator.
Now, the representation of the grand operator can be considered as
\begin{eqnarray}\label{grand}
&{\cal{H}}=\sum_{j=1}^{2N}\sum_{k=1}^{N}\sum_{r=\pm
1}|j,f'_{k_{r}}(j)\rangle\langle\psi '_{(k_{r},j)}|.
\end{eqnarray}
  It is easy to
check the properties ${\cal{H}}|\psi
'_{(k_{r},j)}\rangle=|j,f'_{k_{r}}(j)\rangle$ and also
${\cal{H}}^2=I$.

Again, one can find an explicit form for $\cal{H}$ operator based
on the introduced basic operators.  At first, Bob needs one
operator to disentangle the two particles. We have considered this
operator as a non-local operator which acts conditionally on the
$|l,m\rangle$ state as
\begin{eqnarray}\label{PCS}
{\text{PCS}}|l,m\rangle=\theta(-l)(I\otimes
P_m)|l,m\rangle+\theta(l)|l,m\rangle
\end{eqnarray}
where PCS stands for position controlled swap operator and $\theta
(l)$ is the conventional unit step function.  Elsewhere, we have
proposed a method to realize this gate using four usual CNOT gates
{\cite{Akh,thesis}}.  On the other hand, the same as the
spin-$\frac{1}{2}$ case, Bob can use Hadamard operators in the
position space with the form
\begin{eqnarray}
H_{x_{n}}=\frac{1}{\sqrt{2}}(P_{n}+N_{n}) \label{Hxn}
\end{eqnarray}
where $H_{x_n}$ is a local operator acting on the $\pm n$-th
channels.  But for $N\geq 2$ cases, applying just PCS and
$H_{x_n}$ operators does not produce pure position states in a
disentangled and measurable form.  So, we should introduce a
non-local unitary operator for this mean which acts like
\begin{eqnarray}\label{U}
&\hskip -37mm
U_{(N)}|rl,r'm\rangle=\nonumber\\&{\frac{1}{\sqrt{N}}}\sum_{n=1}^{N}h_{m,m+n-1}|f_{l_{r}}(n),f_{m_{r'}}(n)\rangle_{{\text{mod}}(N)}
\end{eqnarray}
where $l$ and $m$ are positive integers.  Moreover, here,
$h_{m,n}$ is an element of an $N$-dimensional normalized symmetric
Hadamard matrix.  It is straightforward to check that,
$U_{(N)}^2=I$. Considering the works of Barenco {\it et al.}
{\cite{Barenco}} as well as Bremner {\em et al.} {\cite{Nielsen}}
and the Eq.~(\ref{U}), it can be concluded that in construction of
$U_{(N)}$ operator, at most $(N-1)$ successive $L_+$ operators for
each particle and ${\cal{O}}(N)$ parallel PCS gates are certainly
needed. The same as $O_{(k_{r},j)}$ operator, it is also possible
to find an explicit form for $U_{(N)}$ operator based on the basic
gates \cite{thesis}, but its details are omitted here. Now, using
the above mentioned operators, Bob can perform the position BSM in
this way
\begin{eqnarray}\label{Hexplicit}
U_{(N)}(H_{x_1}H_{x_2}\ldots H_{x_n}\otimes
I)\text{PCS}|\psi_{(k_{r},j)}\rangle=|m,n\rangle
\end{eqnarray}
which should be followed by a position measurement on the outgoing
channels.  In Eq.~(\ref{Hexplicit}) the operator $U_{(N)}$ acts on
all the channels, and $m$ as well as $n$ are some unique functions
of $k_{r}$ and $j$. In addition, action of the set of operators in
(\ref{Hexplicit}) are equivalent to the action of the grand
operator (\ref{grand}).

We have seen that, there are $4N^2$ Bell basis and the same number
of different unitary operators $O_{(k_{r},j)}$ for encoding in our
dense coding scheme. This obviously corresponds to encoding $4N^2$
different messages by Alice.  Thus, she can send
$2~{\text{log}}_{2}(2N)$ bits of classical information per
particle to Bob.  Now, he needs $N$ PCS, $2N$ Hadamard and one
$U_{(N)}$ gates to read out the sent classical information during
the BSM.  Since Bob performs one BSM on just the two particles, it
is possible to consider that all PCS and also all Hadamard gates
operate in a parallel form, i.e. {\it{concurrently}}. If the
operation times for the PCS, Hadamard and $U_{(N)}$ gates are
$t_p$, $t_h$ and $t_u$, respectively, the rate of classical
information gain R, defined as sent classical bits of information
per unit time and sent particle, is
\begin{eqnarray}\label{logbitnumber}
R_x=2~{\text{log}}_{2}(2N)/(t_p+t_h+t_u).
\end{eqnarray}
  Similarly, one can calculate $R$ for a dense coding
protocol which works using ${\textsf{N}}$ pairwise entangled
qubits and/or $\textsf{N}$ maximally entangled qubits shared
between two parties {\cite{Bose}}. In the pairwise entangled case,
$2{\textsf{N}}$ classical bits of information are transferred from
Alice to Bob. Meantime, in this case, $\textsf{N}$ separate CNOT
and Hadamard gates are required by Bob to decode the Bell states.
Thus, its $R$ in terms of bits per unit time per sent particle is
\begin{eqnarray}\label{logbitnumber}
R_p=2{\textsf{N}}/{\textsf{N}}^2(t_c+t_h)
\end{eqnarray}
where $t_c$ is the operation time of a CNOT gate.  On the other
hand, in the maximally entangled case, the number of sent
classical bits is $\textsf{N}$. In this case, Bob needs
$({\textsf{N}}-1)$ successive CNOT and one Hadamard gates so that
he gains
\begin{eqnarray}\label{logbitnumber}
R_m={\textsf{N}}/({\textsf{N}}-1)[({\textsf{N}}-1)t_c+t_h]
\end{eqnarray}
bits per time and particle.  Now, if we assume that both $N$ and
${\textsf{N}}$ are very large and all basic gates operate in an
equal time interval, i.e. $t_c\sim t_h\sim t_p/4\sim t_u/N\equiv
t$, then $R_x=2~{\text{log}}_{2}(2N)/Nt$ and
$R_p=R_m=1/{\textsf{N}}t$. Therefore, as is seen, at large $N$'s,
if $N={\textsf{N}}$ is considered, i.e. dimensions of Hilbert
spaces are identical, then our protocol is more efficient than
both the pairwise and the maximally entangled cases with a {\em
logarithmic} factor.

Furthermore, other degrees of freedom such as spin, polarization
and so on can be added to our protocol in order to obtain a more
powerful dense coding.  Here, for instance, we consider that each
particle can also have the spin $S$.  To introduce spin into our
protocol, it is sufficient to assume that the source emits
entangled pair of particles not only with vanishing total momentum
but also with zero total spin. Therefore, the quantum state of the
system would be
\begin{eqnarray}
&|\psi_{1}\rangle_{xs}=\frac{1}{\sqrt{2N(2S+1)}}\sum_{n=1}^{N}[|n,-n\rangle+|-n,n\rangle]
\nonumber\\&\times\sum_{s=0}^{2S}(\pm 1)^{s}|(S-s),-(S-s)\rangle
\end{eqnarray}
which is simply a tensor product of position and spin states of
the system.  Therefore, now Alice is capable of sending
$2~{\text{log}}_{2}[2N(2S+1)]$ bits per particle.

In summary, we have proposed another theoretical extension of
dense coding protocol by using entangled spatial states of the two
particles shared between two parties.  Our construction is based
on using the well known Hadamard matrices for building orthogonal
states, required encoding and decoding operators, and hence is
subject to their intrinsic characteristics.  Furthermore, we have
given a typical proposition for approaching to realization of the
scheme. For this mean, some basic operations have been introduced
and it is shown that whole the scheme can be established based
upon them. By comparing our scheme with some previously proposed
multi-qubit protocols, it is shown that the rate of classical
information gain in our case is better than them with a
logarithmic factor. Also we have shown that considering internal
degrees of freedom, like spin, strengthens the scheme in a
straightforward manner.

 We would like to thank V. Karimipour for his
stimulating discussions on quantum information theory. O.A. and
A.T.R. also appreciate hospitality of the ICTP (Italy) where some
part of this work was completed.
%%%%%%%%%%%%%%%%%%%%%%%%%%%%%%%%%%%%%%%%%%%%%%%%%%%%%%%%%%%%%%%%%%%%%%%%%%%

%%%%%%%%%%%%%%%%%%%%%%%%%%%%%%%%%%%%%%%%%%%%%%%%%%%%%%%%%%%%%%%%%%%%%%%%%%%%%%%%%%%%%%%%%%%%%%%%%%%
\end{document}